# Central role of domain wall depinning for perpendicular magnetization switching driven by spin torque from the spin Hall effect

O. J. Lee,[1] L. Q. Liu,[1] C. F. Pai,[1] H. W. Tseng,[1] Y. Li,[1] D. C. Ralph[1,2] and R. A. Buhrman[1]

[1]Cornell University, Ithaca, NY, 14853, USA

[2]Kavli Institute at Cornell University, Ithaca, NY, 14853, USA

We study deterministic magnetic reversal of a perpendicularly magnetized Co layer in a Co/MgO/Ta nano-square driven by spin Hall torque from an in-plane current flowing in an underlying Pt layer. The rate-limiting step of the switching process is domain-wall (DW) depinning by spin Hall torque via a thermally-assisted mechanism that eventually produces full reversal by domain expansion. An in-plane applied magnetic field collinear with the current is required, with the necessary field scale set by the need to overcome DW chirality imposed by the Dzyaloshinskii-Moriya interaction. Once Joule heating is taken into account the switching current density is quantitatively consistent with a spin Hall angle $\theta_{SH} \approx 0.07$ for 4 nm of Pt.

The electrical manipulation of thin film nanomagnets is currently the focus of strong research interest, in part because this phenomenon offers significant advantages with respect to power consumption and high-speed operation in technological applications including magnetic memory and non-volatile logic. For more than a decade the direct injection of spin-polarized currents has been utilized to induce nanomagnet switching, persistent magnetic oscillation, and magnetic domain wall (DW) motion via the spin-transfer torque (ST) effect[1] in various types of magnetic nanostructures. Recently an alternative method[2-9] for electrical manipulation of magnetic moments has been demonstrated whereby in-plane currents achieve deterministic magnetic reversal in multilayer NM/FM/MO$_x$ samples consisting of a non-magnetic heavy metal (NM) adjacent to a very thin ferromagnet (FM) that is capped with a metal oxide (MO$_x$). This offers a more efficient pathway for nanomagnet control that does not require a magnetic spin polarizer, and that also provides for the separation of the write and read channels in magnetic memory devices[3]. Switching by in-plane currents has been observed for magnetic layers with either perpendicular magnetic anisotropy[2-8] (PMA) or in-plane anisotropy[3,9], via different reversal processes. The switching of PMA layers has the interesting property that it is necessary to apply at least a weak external field $\vec{H}$ with a component ($\geq 100 \pm 50$ Oe for the system studied here) parallel or anti-parallel to the applied current density $\vec{J} = J\hat{x}$, such that the direction of switching is determined by the sign of $\vec{J} \cdot \vec{H}$.[2-4] Here we analyze the microscopic processes in operation during current-driven reversal of PMA samples, and show that for quantitative understanding it is necessary to move beyond a previous macrospin description and consider how the depinning of magnetic domain walls is governed by the combined effects of an in-plane magnetic field and the torque induced by the in-plane current.

The current-driven torque at work within NM/FM/MO$_x$ structures with PMA has been attributed to two different mechanisms[5,10-12]; a Rashba effect[2,6-8,13] (RE) within the FM layer and a spin-Hall effect[3,4,9,14-16] (SHE) within the NM layer. In the RE case it has been proposed that due the dissimilar interfaces (NM/FM or FM/MO$_x$) of the FM layer there can be a substantial intrinsic interfacial electric field in the vertical ($\hat{z}$) direction, and spin-orbit interactions (SOIs) generated by this field can cause an electrical current density $\vec{J} = J\hat{x}$ to induce a Rashba effective magnetic field $\vec{H}_R = H_R\hat{y} \propto \hat{z} \times \vec{J}$ in the in-plane direction transverse to the current, and also potentially an equivalent magnetic field component in the $\hat{m} \times \hat{y}$ direction that can drive magnet reversal (here $\hat{m}$ is the magnetic orientation direction of the FM). The other proposed mechanism is that there is a substantial SHE whereby $\vec{J} = J\hat{x}$ in the NM generates a transverse spin current ($\hat{\sigma} // \hat{y}$) via SOIs. The absorption of this in-plane polarized spin current at the NM/FM interface exerts a ST per unit moment $\vec{\tau}_{ST} = \gamma H_{SH} \hat{m} \times (\hat{y} \times \hat{m})$ on the FM. Although $\vec{\tau}_{ST}$ has an anti-damping effect on a FM moment when the spin-polarization ($\hat{\sigma}$) has a component collinear to $\hat{m}$, when ($\hat{\sigma} // \hat{y}$) is orthogonal to a substantial component of $\hat{m}$, as for example in the out-of-plane magnetized case, $\hat{m} \approx \hat{z}$, the ST can also be understood as equivalent to a longitudinal equivalent magnetic field $H_{SH} \hat{m} \times \hat{y}$ acting on the FM moment whose strength is linearly dependent on $J$. Previously[4] it has been argued that, given values of the strength of the SHE in Pt measured independently, the SHE mechanism by itself is sufficient to explain the switching in Pt/FM/MO$_x$ structures with PMA, and we will provide additional evidence for that conclusion below. However the process of domain wall depinning that we will describe would apply to any current-driven equivalent field with $\hat{m} \times \hat{y}$ symmetry, regardless of origin.

A previous analysis of reversal of PMA magnetic layers by in-plane currents employed a simple macrospin picture of magnetic dynamics.[4] This gives reasonable agreement with the measured switching phase diagram (SPD) assuming the action of an equivalent field with $H_{SH}\hat{m} \times \hat{y}$ symmetry, as long as the measured change of the coercive field as a function of Joule heating is accounted for by hand. However, a macrospin description is clearly inadequate for providing accurate quantitative understanding of the reversal process. First, in actual macrospin switching, the current-induced equivalent field needed to reverse a sample with PMA should be $H_{SH} \approx H_k^{eff}/2$,[4] where $H_k^{eff}$ is the effective anisotropy field of the PMA layer which is generally very large, e.g. $H_k^{eff} = 2.8$ kOe in ref. [4], but switching actually occurs for much smaller current-induced equivalent fields, $H_{SH} \approx 300$ Oe.[4] This difference has already been ascribed to reversal by a process of nucleating reversed domains and propagating domain walls (instead of macrospin-like coherent rotation and switching) that lowers the coercive field well below $H_k^{eff}/2$, but as of yet there is no microscopic picture that provides a framework for understanding the switching process quantitatively. A second inadequacy of the macrospin model is that there is no explanation for the scale of the magnetic field that must be applied collinear with the current direction in order for the current-driven switching to proceed.

Here we report an experimental study of the current-induced switching of perpendicularly-magnetized Co ($\perp$Co) thin-film nano-squares (100's of nm on a side) formed on a Pt microstrip as a function of in-plane bias current and magnetic field. Through measurement of the critical currents for switching and activation energy barriers $E_a$ we confirm that reversal occurs by the nucleation of reversed domains much smaller than the device size followed by a thermally-assisted DW depinning process that results in the complete reversal of the entire Co by DW propagation. We identify the rate-limiting step of reversal as spin-Hall-torque-driven DW depinning. The role of the in-plane magnetic field is to turn the in-plane orientation of the magnetic moments within the domain wall to have a significant component parallel to the current flow, thereby allowing the torque from the spin Hall effect to produce a perpendicular equivalent field that can expand a reversed domain in all lateral directions. This model provides a quantitative explanation why, once Joule heating is taken into account, only a relatively small ST equivalent field $H_{SH,z}$ is required to drive full reversal, and it also explains the scale of the required in-plane applied magnetic field. We find that spin Hall torque with a strength corresponding to a spin Hall angle of $\theta_{SH} \approx 0.07$ for a 4 nm Pt layer provides a quantitative description for all of our reversal data.

*Sample fabrication and measurement:* For this study we fabricated cross-bar devices for Hall-effect measurements (see Fig. 1a) from a thin film multilayer consisting of, from bottom to top, Pt(4)/Co(0.8-1.0)/MgO(2)/Ta(2) (thicknesses in nm), deposited on thermally oxidized Si substrate by DC/RF magnetron sputtering at room temperature. The base pressure was $< 2 \times 10^{-8}$ Torr and the deposition rates were low (< 0.3 Å/s). The thin Ta capping layer was employed to protect the MgO from degradation due to water vapor exposure during processing and storage. Since thin Ta metal layers are highly resistive even if not fully oxidized by the exposure to the atmosphere any current-shunting effect of the Ta layer was negligible in our experiments. We used e-beam lithography and ion milling to define the current channel and the detection channel, varying these dimensions from 200 nm to 1000 nm. We then used a second stage of aligned e-beam lithography and ion milling to pattern the Co layers into square shapes, with the etching stopped as close as possible at the Pt/Co interface by the use of mass-spectroscopy monitoring of the sputtering process. Finally we evaporated Ti(5 nm)/Au(100 nm) onto contact regions defined by photo-lithography. We will report in detail on the behavior of one particular device with $l$ (detection channel length) = $w$ (current channel width) ≈ 300 nm, with the Co layer dimensions of 300 nm × 300 nm and $t_{Co} \approx 0.9$ nm. The behavior was quite similar for all of the devices studied (> 30). After fabrication the devices were annealed under high vacuum (< $5 \times 10^{-7}$ Torr) at 320 °C for 1 hour to enhance the PMA of the Co. For the results reported below the perpendicular component of the Co-layer magnetization was monitored by applying a DC current ($I_{DC}$) through the current channel (see Fig. 1a) and measuring the extraordinary Hall resistance $R_H = V_H/I_{DC}$ that developed across the orthogonal detection channel due to that part of the bias current that flowed through the Co. Alternatively we also measured the differential extraordinary Hall resistance ($r_H = dV_H/dI_{AC}$) that resulted from a small AC current ($I_{AC} \approx 20$ µA) flowing through the current channel. We utilized the current-channel resistance ($R_c$) as a sensor for the increase in the temperature $T(I)$ of the device due to Joule heating.

The inset of Fig. 1b shows the hysteresis loop obtained by measuring $r_H$ w.r.t an applied out-of-plane field ($H_z$) for $I_{DC} = 0$ mA, indicating that the device has a good PMA with an out-of-plane switching field $H_p \approx 360$ Oe. Application of an in-plane field ($H_x$) (not shown) indicates that $H_k^{eff} \approx 4$ kOe as determined from the fitting to the hard-axis magnetic field dependence of $r_H$.

*Magnetic reversal behavior of Co nano-squares:* To quantify the effect of the current induced torque in reversing the perpendicularly-magnetized Co ($\perp$Co) layer it is first necessary to understand the basic nature of the reversal process. If the $\perp$Co nano-square followed Stoner-Wohlfarth (SW) macrospin behavior[17], the thermally-activated switching field ($H_p$) should be close to $H_k^{eff}$ in any finite measurement time because the thermal stability factor $\Delta = E_{a,ideal}/k_B T_o$ for this reversal would be very large, since $E_{a,ideal} \approx M_s H_k^{eff} V / 2 \approx 100$ eV. (Here $V = l\,w\,t_{Co}$ is the Co nano-square volume and $4\pi M_s \approx 13$ kOe.) However the observed $H_p$ is much smaller and varies as a function of the measurement time ($t_m$) as illustrated in Fig. 1b and as discussed below. In addition, in the SW case the reversal field $H_p(\theta)$ should vary as $H_p(\theta)/H_p(0) = 1/\left(\cos^{2/3}\theta + \sin^{2/3}\theta\right)^{3/2}$, where $\theta$ is the tilt angle of the applied field from the out-of-plane easy axis (see inset Fig. 1c.). As shown in Fig. 1c we find instead $H_p(\theta)/H_p(0) \approx 1/\cos\theta$, which is as predicted[18,19] for the case of domain wall depinning, with some deviation from that behavior as $\theta$ approaches 90º. We conclude that magnetic reversal occurs via first the nucleation of one or more reversed domains in the Co followed by the thermally activated depinning of DWs[20,21] that completes the reversal, with the high field departure from a $1/\cos\theta$ dependence most likely due to coherent rotation of the magnetization vector in the pinned domain when the in-plane hard-axis field component is sufficiently strong[19]. Reversal via depinning is also consistent with the magnetic hysteresis loop $r_H$ vs. $H_z$ shown in the inset of Fig. 1b which indicates a small, $\leq 5\%$, change in $r_H$ ($\propto M_z$) before the full reversal that we attribute to the time-averaged presence of one or more small pinned domains[22] within the Co nano-square for applied fields smaller than the depinning field $H_p$.

To determine the depinning field ($H_{p,0}^z$) in the absence of thermal fluctuations for out-of-plane fields and the activation energy barrier ($E_p$) for the depinning, we performed a ramp-rate measurement at $T_o = 295$ K, measuring the average switching field $\langle H_p^z \rangle$ ($\equiv \langle H_p \rangle$) as a function of $t_m$ (see Fig. 1b). We obtained the thermal stability factor $\Delta = E_p/k_B T_o = 38 \pm 8$ ($E_p = 0.95 \pm 0.2$ eV) and $H_{p,0}^z = 900 \pm 200$ Oe from a fit to the standard model[21] for thermally assisted depinning:

$$\langle H_p \rangle = H_{p,0}^z \left\{ 1 - \left[ \frac{k_B T_0}{E_p} \ln\left(\frac{f_0 t_m}{\ln 2}\right) \right] \right\}, \tag{1}$$

where $f_o$ is the characteristic fluctuation attempt frequency (assumed here to be $f_o = 10$ GHz).

While the origin of this pinning is not critical to the analysis of the current assisted switching behavior that is the main focus of this work, we tentatively attribute it to spatial variations in the effective anisotropy field $H_k^{eff}$, as have been examined recently for domain wall pinning in PMA nanowires[20]. In the case of our ⊥Co, spatial variations in $H_k^{eff}$ could arise from, for example, grain-to-grain variations in the interfacial anisotropy energy density $K_i$ and/or variations in the Co thickness $t_{Co}$ since $H_k^{eff} = 2K_i / M_s t_{Co} - 4\pi M_s$. We also note that $M_s$ can vary strongly with $t_{Co}$ in this ultra-thin film regime[23]. If this attribution is correct then the pinning field $H_{p,0}^z \approx 900$ Oe ($= \Delta H_k^{eff} \equiv H_{k,\max}^{eff} - H_{k,\min}^{eff}$ according to ref. [20]) indicates that since $H_{k,\max}^{eff} \approx 4$ kOe there is an ~25% variation in the anisotropy field between the value $H_{k,\min}^{eff}$ averaged over the minimum volume $V_s$ required to support a previously nucleated sub-volume domain and that of the surrounding area $H_{k,\max}^{eff}$ as illustrated schematically in Fig. 1d.

We can estimate the size of the minimum volume $V_s$ that is required to support nucleation of a reversed domain by noting that for thermally activated depinning to be the rate limiting step in the reversal, it is necessary[20,21] that the activation energy for domain nucleation $E_n$ must be < $E_p$. Since $E_n = K_{eff,\min} V_s$, where $K_{eff,\min} = H_{k,\min}^{eff} M_s / 2$, this requires that the diameter of $V_s$ be < 37 nm, much smaller than the sample. This value is also quite compatible with the requirement that $V_s \geq \pi \delta_{dw}^2 t_{Co}$ where $\delta_{dw} = (A_{ex} / K_{eff})^{1/2}$ is the domain wall thickness and $A_{ex}$ is the exchange-stiffness ($\approx 1.6 \times 10^6$ erg/cm), which from other work[24] results in $\delta_{dw} \approx 9$ nm. Nucleation of similarly-small domains has been recently proposed[25,26] to explain why in MgO magnetic tunnel junctions that incorporate very thin CoFeB electrodes with PMA the thermal stability is almost invariant with junction area once the lateral dimensions are > 40 nm.

*Current assisted switching*: We studied the ability of a DC current density $J$ flowing through the Pt channel to modify the thermally activated magnetic switching of the ⊥Co nano-square for the cases (a) where the external field $H_z$ is applied in the out-of-plane easy axis direction, and (b) where the field is applied in-plane, both in the direction perpendicular to the current flow, $H_y$, and collinear with the current, $H_x$. If the magnetic reversal of the ⊥Co occurs via DW depinning of a previously-nucleated domain, then the mechanism by which $J$ assists the magnetic reversal should be describable in terms of the effect on the energy barrier for depinning due to any SHE-generated out-of-plane equivalent field ($H_{SH,z}$) acting on the domain wall magnetization. In this case we would expect from Eq. (1) that the stability factor for the depinning ($\Delta = E_p/k_B T$) should become, after taking into account both the SHE equivalent field and Joule heating,

$$\Delta^* = \left(\frac{E_p(J)}{k_B(T_0+\kappa J^2)}\right)\left(1-\frac{H_z+H_{SH,z}(J)}{H_{p,0}^z(J)}\right), \qquad (2)$$

where $H_z$ is any out-of-plane applied field, $T_o = 295$ K, $E_p(J)$ is the depinning energy barrier as a function of $J$ (or increased $T$) and $H_{p,0}^z(J)$ is the effective pinning field in the absence of thermal fluctuations as a function of $J$ (or $T$).

Since for the SHE $\vec{H}_{SH} = H_{SH}\hat{m}\times\hat{y}$, the out-of-plane component of the spin Hall equivalent field experienced by the domain wall depends not only on the strength and direction of the current density $J$ but also on the orientation $\hat{m}$ of the magnetization within the domain wall. More precisely, the vertical component of the spin Hall equivalent field is $H_{SH,z} = H_{SH}m_x$, where $m_x$ is the magnetization component collinear with the current. Thus if $m_x$ always had the same sign within the domain wall we would expect that a fixed $J$ would either enhance or decrease the switching field $H_p$ required for a reversal, or equivalently for a fixed field bias one direction of $J$ would effectively increase the total $H_z$, resulting in a reversal, while the opposite direction would decrease $H_z$, making a thermally activated transition less likely until Joule heating became sufficiently strong. This is not the case when the applied field is out of plane. In Fig. 2a we show the result of a measurement where the ⊥Co was initially set to $\hat{m}_z = +\hat{z}$ and $H_z$ = -200 Oe was applied. Then $J$ was swept back and forth, beginning in either a positive or a negative initial direction. In all cases, we observed that in the first sweep only there was an abrupt transition in $r_H$ at essentially the same switching current density $|J_s| \approx 34$ MA/cm$^2$ ($|I_s| \approx$ 0.5 mA) regardless of current polarity. We conclude that Joule heating initiates the reversal during the first sweep to the stable low-energy state where $m_z$ is aligned with $H_z$. Note that at higher currents, $|J| \approx 64$ MA/cm$^2$ ($|I_p| \approx 0.94 \pm 0.02$ mA), there are sharp changes (a peak or dip) in $r_H$, above which it quickly converges to zero. Since the value of $J$ at which these abrupt changes occur is independent of the external field orientation, we attribute this latter behavior to the loss of PMA in the Co due to heating[27].

In Fig. 2c we plot the SPD which shows the combinations of bias current density $J$ and $H_z$ that result in transitions from the bistable region where the ⊥Co moment can be either up or down, as determined by an initializing field bias step, to the regions where the current and field values are such that the moment is either uniquely up or down. The SPD is symmetrical about the $H_z$ axis and also about the $J$ axis. The straightforward conclusion is that for $H_z$ field biases, the only significant effect of $J$ is Joule heating, which promotes the thermally activated depinning in the same way for both current directions. When the applied field is in-plane but transverse to the current flow, $\vec{H} = H_y\hat{y}$, the result from a current ramp is essentially the same; the only transition caused by the current is the apparent loss of PMA due to heating at $|J| \sim 65$ MA/cm$^2$.

When an in-plane magnetic field is applied in the direction collinear with the current flow the behavior is quite different, provided $\vec{H} = H_x \hat{x} \geq 100$ Oe, consistent with previous observations[2,4,28,29] of deterministic switching with Pt/Co/AlO$_x$, Pt(thick)/Co/Pt(thin), and Pt/CoFe PMA structures. We obtain, as illustrated in Fig. 2d-e for $H_x = \pm 300$ Oe, clear current-induced deterministic switching of the ⊥ Co nano-square. The sudden reversals in $r_H$ and sharp jumps in $R_H$ at the current value $|J_s| \sim 44$ MA/cm$^2$ ($|I_s| \sim 0.65 \pm 0.02$ mA), are indicative of switching of the ⊥ Co from $m_z \approx +\hat{z} \to -\hat{z}$ when $\vec{J} \cdot \vec{H} > 0$ and from $m_z \approx -\hat{z} \to +\hat{z}$ at essentially the same current magnitude when $\vec{J} \cdot \vec{H} < 0$. The SPD for applied fields in the $x$ direction is shown in Fig. 2f. Next we consider what sets the scale of the applied field $H_x$ that is required to achieve this reversible, deterministic switching.

Studies of current-driven domain wall motion in NM/FM bilayers[28-34] have shown that for a SHE torque to be effective in displacing a DW in a PMA material that DW has to be a Neel wall (NW) rather than a Bloch wall (BW). While BW's are generally more energetically favorable for extended domains in thin magnetic layers due to the demagnetization energy of the NW, this difference becomes small in the ultra-thin film limit and can reverse in patterned nanostructures[29,32]. Moreover, work has shown that at the interfaces of very thin NM/FM/MO$_x$ layers there can be a strong Dzyaloshinskii-Moriya interaction (DMI) such that in a patterned nanowire a Neel Wall (NW) with a fixed chirality[29,32] is energetically favored as schematically illustrated in Fig. 1d. This is key to the successful interpretation of experiments[29,32] in which a DW can be rapidly displaced along such a nanowire wire by a bias current, with the direction of the displacement dependent upon the sign of the spin Hall angle ($\theta_{SH}$) of the high Z NM (such as Pt or Ta).

For the magnetic reversal that is of interest here the requirement is that SHE torque assist the *expansion* of a domain in all directions rather than the displacement of a domain wall in one direction. For the former to occur it is necessary that the in-plane magnetization at the center of the domain wall on all sides of an enclosed domain have a component that is collinear with the direction of current flow and that the sign of this collinear component be the same throughout the DW (Fig. 3c). This requires an applied field $|H_x| > 0$ and if there is a substantial DMI then $H_x$ must be strong enough to break any chirality in the DW imposed by the DMI (left-handed for the Pt/Co case). Qualitatively this is consistent with the observation that reliable reversal is not obtained for our ⊥Co for $|H_x| < 100$ Oe, *e.g.*, see Fig. 3b for $H_x = 50$ Oe.

We can employ this model to obtain quantitative estimates for the strength of the spin Hall torque and the DMI in our sample, using the ⊥Co SPD, as shown in Fig. 2f. In our analysis we use Eq. 2 with $H_z = 0$ and also the result that for a Neel DW with the in-plane magnetic orientation of the pinned DW fully aligned with $H_x$ the torque exerted on the wall from the SHE, when averaged over the thickness of the wall, is equivalent to a magnetic field

$$H_{SH,z} = 1/\pi \int_0^\pi H_{SH} \sin\varphi \, d\varphi = (2/\pi) H_{SH}$$, where $\varphi$ is the local orientation of the DW magnetization relative to $\hat{z}$. (See Fig. 3c.)

To estimate the strength of the SHE torque we first have to quantify the effects of Joule heating, for which we used the $\langle J_s \rangle$ vs. $H_z$ SPD as measured with $H_x = 0$; the case where we have concluded there is no effect from $H_{SH}$ because the net $H_{SH,z}$ over the DW is approximately zero for a chiral domain wall (Fig. 1d). We calibrated the channel resistance $R_c(T)$ by heating the substrate externally, and then separately measured $R_c$ as a function of $|I|$ applied to the current channel, from which we obtained $T(J) \approx T_o + \kappa J^2$ with $\kappa = 0.05$ K cm$^4$/MA$^2$, quite similar to the heating rate obtained recently from a study of similar Pt/Co multilayers[35]. From this we estimate that $T \approx 390$ K at $|J_s| = 44$ MA/cm$^2$ ($|I_s| = 0.65 \pm 0.02$ mA), the point where the SHE switching shown in Fig. 2d-e occurs, and $T \approx 495$ K at $|J| = 64$ MA/cm$^2$ ($|I| = 0.94 \pm 0.02$ mA), the point where the PMA begins to decrease rapidly as shown Fig. 2a-b and d-e. Next we estimated the Curie temperature of the $\perp$ Co, $T_c \sim (583 \pm 23)$ K, from a fit to the empirical relationship[36] $\Delta R_H(T(J)) = R_{H0}(1 - (T(J)/T_c)^\alpha)^\beta$, as shown in Fig. 3a, where $R_H(I) = \Delta V_H(I)/2I$ is the maximum Hall-resistance (for $|m_z| \approx 1$) at a given $I$, by measuring the difference of $V_H$ at large $H_z = \pm 1.5$ kOe, under the assumption[23] that $R_H(T(J)) \propto M_s(T)$. As a check, this estimated $T_c$ is very close to the previously reported value ($\sim 600$ K) for a similar thickness of Co sandwiched between two Pt layers[24,37].

To model the effect of heating on the thermally activated depinning, we made the assumption that the depinning energy $E_p(J) \propto M_S(J) H^z_{p,0}(J)$. To the degree that the same measurement time $t_m$ is used to obtain the SPD data points (Fig. 2c) the stability factor $\Delta^*$ is a constant along the phase boundary and we can then employ Eq. 2 to obtain $H^z_{p,0}(J)$, using the direct determination of $M_s(J)$ from the $R_H(T(J))$ measurement. In Fig. 4a we plot the normalized results, $\eta(J) \equiv H^z_{p,0}(J)/H^z_{p,0}(0)$ and $\chi(J) \equiv E_p(J)/E_p(0)$, along with $\xi(J) \equiv M_s(J)/M_s(0)$. While the increase in the depinning field ($\eta(J)$) with $T(J)$ may appear counter-intuitive it is consistent with the reduction in $M_s$ since $H^z_{p,0} \propto \Delta H^{eff}_k$ and

$$H^{eff}_k = 2K_i/M_s t_{Co} - 4\pi M_s.$$

We used this approximate variation of the depinning energy and pinning field with bias together with the $\langle J_s \rangle$ vs. $H_x$ SPD as measured for $H_z = 0$ (Fig. 2f) and Eq. 2 to determine $H_{SH,z}/J$ for the different bias fields $H_x$. The results, plotted in Fig. 4b, vary from ~ 0.7 Oe cm²/MA to ~ 2.5 Oe cm²/MA as $H_x$ ranges from 100 Oe to 600 Oe. We tentatively attribute this variation in $H_{SH,z}$ to an increasingly better alignment of the in-plane magnetization of the DW as a function of the applied field, as $H_x$ increases from the value where it first begins to alter the DW chirality produced by the DMI effective field ($H_{DMI}$) at ~ 100 Oe up to the point where $H_x \gg H_{DMI}$ so that $\hat{m}_{DW} \approx \hat{x}$ and $H_{SH,z}$ is maximized for a fixed $J$. The value of $H_{SH,z}/J$ in the higher field regime ($(2/\pi)H_{SH}/J \approx 2.4 \pm 0.5$ Oe cm²/MA after accounting for the reduction in $M_S$ by Joule heating) corresponds to a spin Hall angle $\theta_{SH} = (H_{SH}/J)(2eM_s t_{Co})/\hbar \approx 0.07$. This is in good quantitative agreement with the value expected from the SHE for a 4 nm Pt layer as previously reported from ST ferromagnetic resonance measurements[16]. If $H_x = 600$ Oe is the approximate point where the in-plane external field fully overcomes the DMI field, $H_{DMI} = D/(M_s \delta_{dw})$, this result indicates $D \approx -0.54$ erg/cm², similar to the value reported from Pt/CoFe domain wall experiments[29].

In summary, we have studied magnetic reversal driven by spin Hall torque for perpendicularly-magnetized Co/MgO/Ta samples on a Pt nanostrip. We have found that the rate-limiting step in the reversal of the $\perp$Co is thermally-assisted depinning under the influence of a SHE-induced equivalent field. This drives expansion of a reversed domain to achieve full reversal of the nano-square. For the SHE to be effective in causing a deterministic reversal by DW depinning it is required that there be an in-plane applied field $H_x$ sufficient to overcome the DW chirality imposed by the Dzyaloshinskii-Moriya interaction so that for the magnetization within the DW $\hat{m}\cdot\vec{J}$ has a uniform sign around the majority of the DW. Our results indicate that for our system the required applied field is approximately 10-25% of $H_{DMI}$. The current-induced equivalent field from the spin Hall effect ($H_{SH,z}/J \approx 2.4 \pm 0.5$ Oe cm$^2$/MA) that we estimate from the thermally activated switching measurements is fully consistent with the expected value (($2/\pi$) $H_{SH}/J \approx 2.3$ Oe cm$^2$/MA) for the case where the underlying Pt layer has a spin Hall angle 0.07.

This research was supported in part by ONR, ARO and NSF. This work was performed in part at the Cornell NanoScale Facility, a node of the National Nanotechnology Infrastructure Network (NNIN) which is supported by the NSF (ECS-0335765), and benefited from use of the facilities of the Cornell Center for Materials Research, which is supported by the NSF/MRSEC program (DMR-1120296).

**Figure Captions**

**Fig. 1** (Color Online) **(a)** Schematic of the cross-bar device structure for Hall-effect measurements. A Co/MgO/Ta nano-square is patterned at the center of a Pt cross-bar. **(b)** Measured average switching field $\langle |H_p| \rangle$ (for an out-of-plane field) for a 300 nm × 300 nm Co nano-square as a function of the measurement time ($t_m$) at room temperature. Inset: Example of a hysteresis loop, $r_H = dV_H/dI$ as a function of $H_z$ at $I_{DC} = 0$. **(c)** Measured switching field $H_p(\theta)$, normalized by $H_p (\theta = 0º)$, as a function of a tilt angle ($\theta$) from the easy axis. The solid line shows the predicted behavior for reversal via domain wall depinning. The dotted line shows the ideal Stoner-Wohlfarth prediction for a single domain nanomagnet. **(d)** Schematic of reversed domain having a domain wall with a fixed chirality (left-handed) due to the Dzyaloshinskii-Moriya interaction. (Red dots = down moments, blue dots = up.) The direction of the out-of-plane component of the equivalent spin Hall field, $H_{SH,z}$ on the domain wall magnetization is indicated schematically for the right and left regions of the wall where $\hat{m} = \pm \hat{m}_x$, respectively.

**Fig. 2** (Color Online) **(a-b)** Examples of hysteresis curves for magnetic switching, from differential Hall measurements, for an out-of-plane external applied field $H_z = \pm 200$ Oe. **(c)** Switching phase diagram of the $\perp$Co showing the average switching current as a function of $H_z$. **(d-e)** Examples of the current-induced deterministic switching of the $\perp$Co under an in-plane external $H_x = \pm 300$ Oe collinear with the current. **(f)** Switching phase diagram of the $\perp$Co showing the average switching current as a function of $H_x$.

**Fig. 3** (Color Online) **(a)** The estimated Curie temperature ($T_c$) of the $\perp$Co ≈ 583 ± 23

K was determined by fitting to $\Delta R_H(T(J)) = R_{H0}(1-(T(J)/T_c)^\alpha)^\beta$ where $R_H(I) = \Delta V_H(I)/2I$ is the maximum Hall-resistance (for $|m_z| \approx 1$) at a given value of $I$ for large $H_z = \pm 1.5$ kOe. (The fit parameters are $R_{H0} \approx 0.96\ \Omega$, $\alpha \approx 0.59$ and $\beta \approx 0.69$.) We assume that $R_H(T(I))$ is linearly proportional to $M_s(T)$. **(b)** Deterministic current-driven switching is absent in the $\perp$ Co nano-square for $H_x = 50$ Oe. There is no magnetic reversal unless the heating is sufficient to destroy the PMA, then upon cooling the PMA is restored with seemingly random orientation $m_z = \pm 1$. **(c)** Left: Schematic of a domain wall structure in which the chirality favored by the DMI (Fig. 1d) is eliminated by a large external field $H_x$. Magnetic reversal occurs when $m_x > 0$ throughout the majority of the domain wall so that the equivalent out-of-plane field from SHE is strong enough to drive *expansion* of a domain in all lateral directions. The direction of the out-of-plane component of the equivalent spin Hall field, $H_{SH,z}$ on the domain wall magnetization is indicated schematically for the left, right, top and bottom regions of the domain wall where $\hat{m} = \hat{m}_x$ in all cases due to the strong $H_x$. Right top: A cross-sectional schematic of the device structure where an electrical current density $J_x$ in the Pt generates a transverse spin current $J_s$ that exerts an effective spin Hall field via spin transfer torque on the spatially varying magnetization of the Co. Right bottom: Polar representation of the spin Hall generated equivalent field, small light (red) arrows, as the function of the Co magnetization direction, large dark (blue) arrows.

**Fig. 4** (Color Online) Estimate for the current induced spin-Hall equivalent field per unit current density ($H_{SH,z}/J$) considering the effects of heating on the magnetic system. **(a)** Estimated effects of Joule heating: $\eta(J)$ is the normalized reduction in $H^z_{p,0}$, $\xi(J)$ is normalized reduction in $M_s$, and $\chi(J)$ is the normalized reduction in $E_p$. **(b)** Values of $H_{SH,z}/J$ obtained from analysis of the switching phase diagram (Fig. 2f).

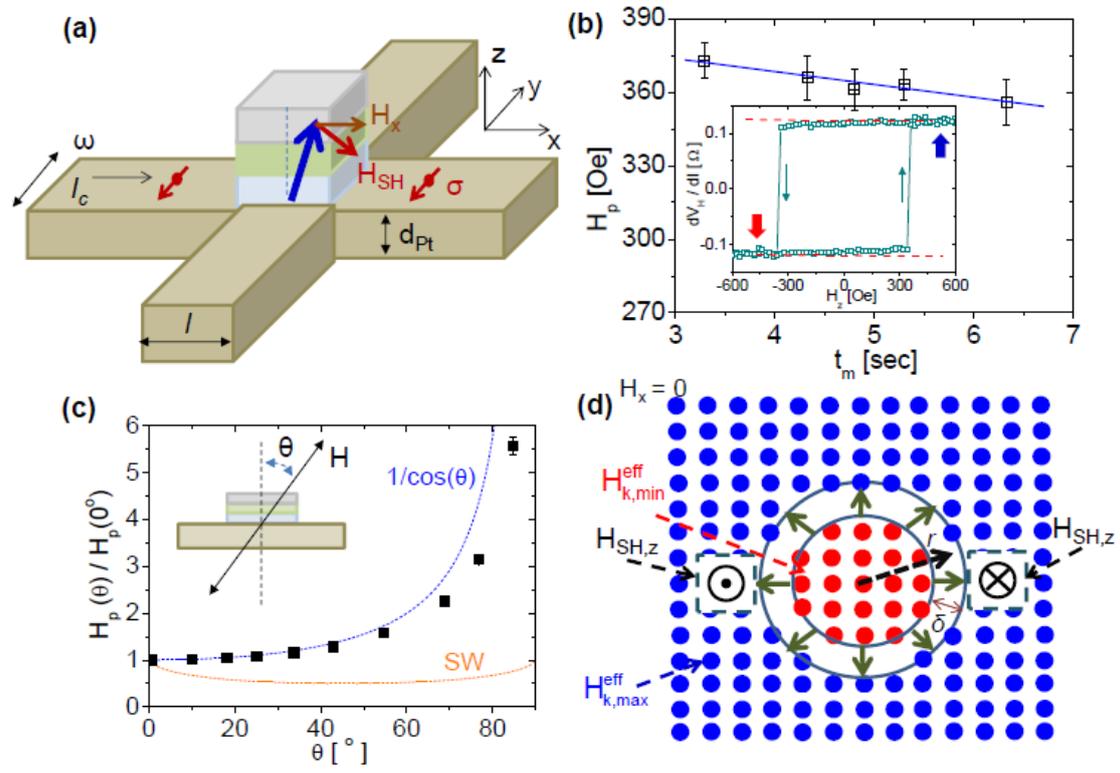

**Figure 1**

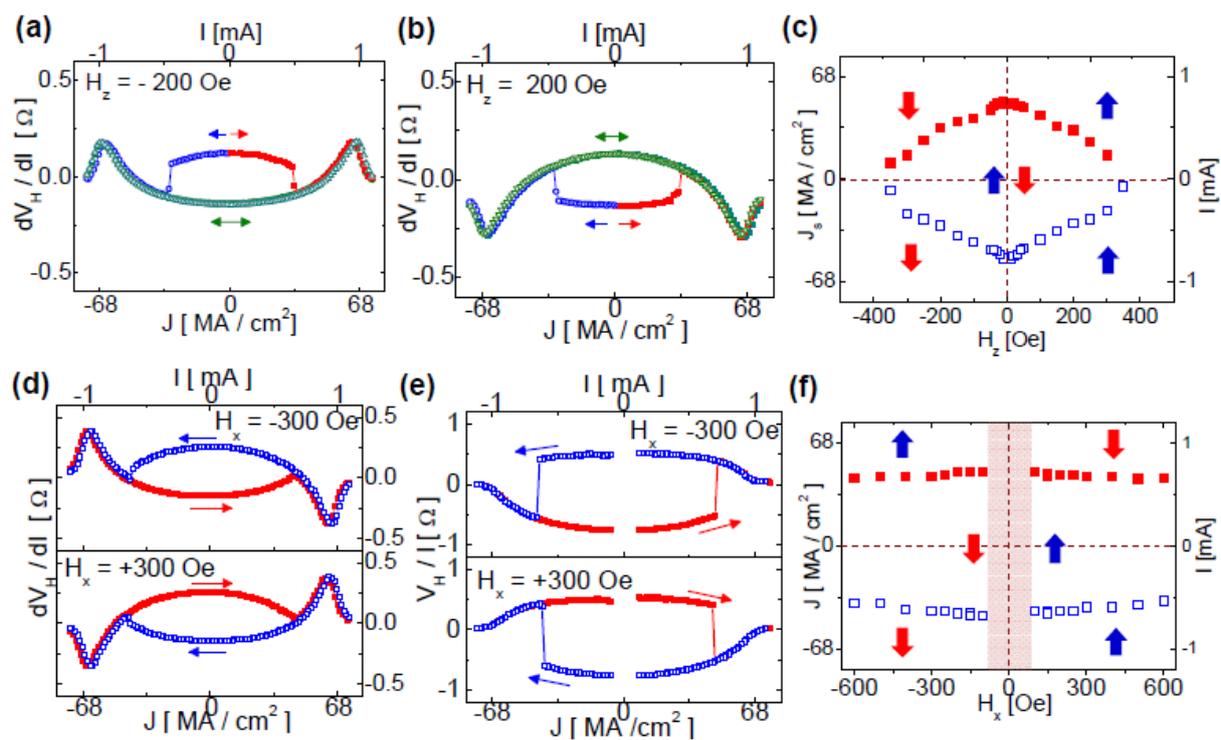

**Figure 2**

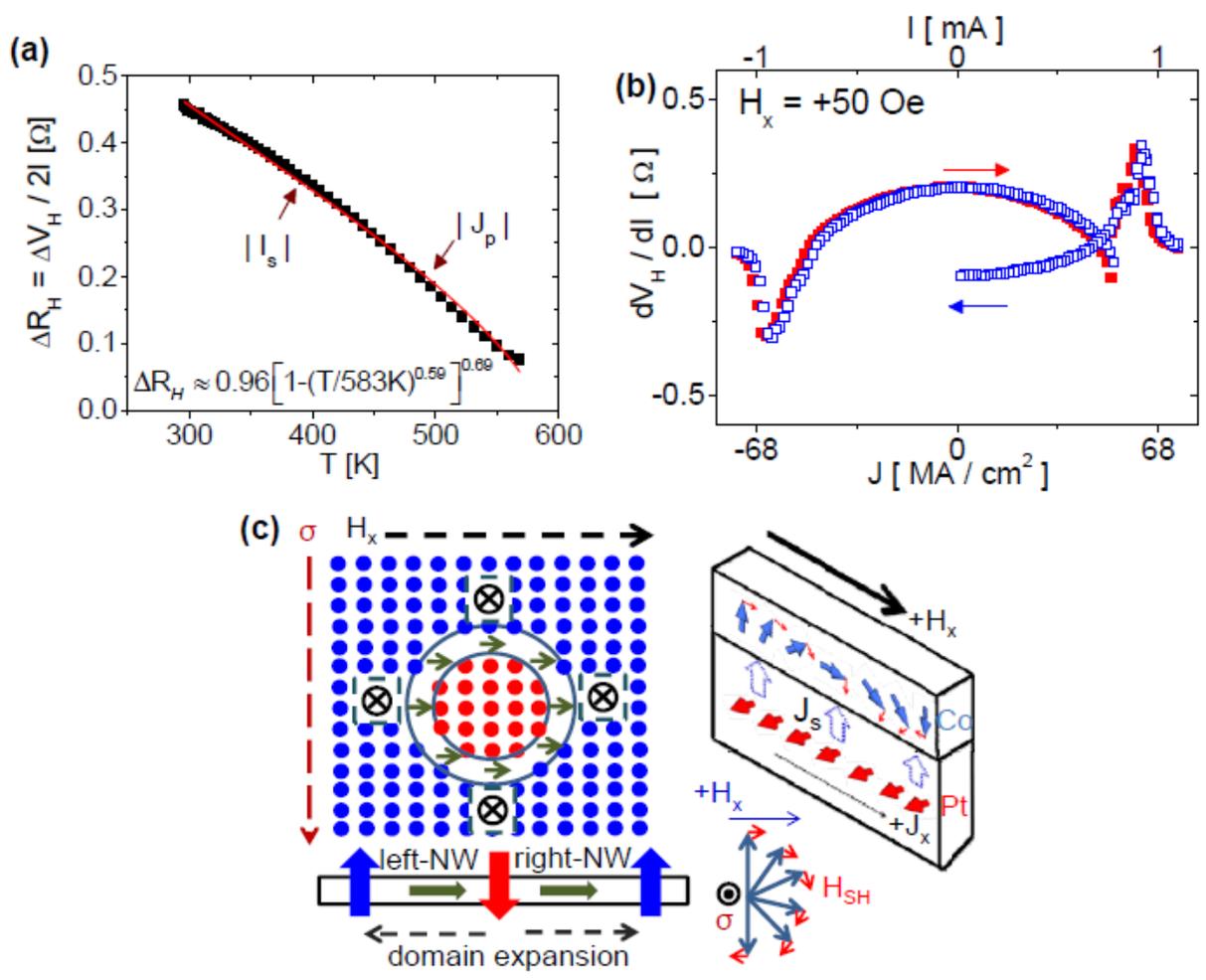

**Figure 3**

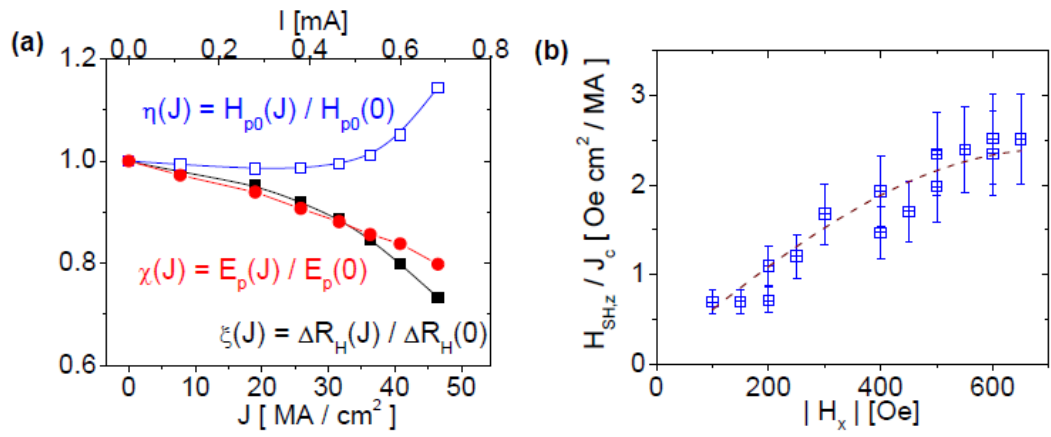

**Figure 4**